\documentclass[usegraphicx,useAMS,usenatbib]{mn2e}
\title[Holographic Imaging]{Holographic Imaging of Crowded Fields:
  High Angular Resolution Imaging with Excellent Quality at Very Low
  Cost} 
\author[R. Sch\"odel, et al.]{R.
  Sch\"odel$^{1}$\thanks{E-mail: rainer@iaa.es}, S. Yelda$^{2}$,
  A. Ghez$^{2}$, J. H. Girard$^{3}$, L. Labadie$^{4}$,
  \newauthor R. Rebolo$^{5}$, A. P\'erez-Garrido$^{6}$, M.R. Morris$^{2}$\\
  $^{1}$Instituto de Astrof\'isica de Andaluc\'ia (CSIC), Glorieta de
  la
  Astronom\'ia S/N, 18008 Granada, Spain\\
  $^{2}$Department of Physics and Astronomy, UCLA, Los Angeles, CA 90095-1547, USA\\
  $^{3}$European Southern Observatory (ESO), Casilla 19001, Vitacura,
  Santiago, Chile\\
  $^{4}$I.Physikalisches Institut, Universit\"at zu K\"oln,
  Z\"ulpicher
  Str. 77, 50937K\"oln, Germany\\
  $^{5}$Instituto de Astrof\'isica de Canarias, C/ V\'ia L\'actea
  s/n, La Laguna, Tenerife E-38200, Spain\\
  $^{6}$Universidad Polit\'ecnica de Cartagena, Campus
  Muralla del Mar, Cartagena, E-30202, Spain\\
}

\begin{document}

\date{}

\pagerange{\pageref{firstpage}--\pageref{lastpage}} \pubyear{2002}

\maketitle

\label{firstpage}

\begin{abstract}
  We present a method for speckle holography that is optimised for
  crowded fields. Its two key features are an iterative improvement of
  the instantaneous Point Spread Functions (PSFs) extracted from each speckle frame and the
  (optional) simultaneous use of multiple reference stars. In this
  way, high signal-to-noise and accuracy can be achieved on the PSF
  for each short exposure, which results in sensitive, high-Strehl
  reconstructed images.  We have tested our method with different
  instruments, on a range of targets, and from the N[10\,$\mu$m]- to the
  I[0.9\,$\mu$m]-band. In terms of PSF cosmetics, stability and Strehl ratio,
  holographic imaging can be equal, and even superior, to the
  capabilities of currently available Adaptive Optics (AO) systems, particularly at
  short near-infrared to optical wavelengths. It outperforms lucky
  imaging because it makes use of the entire PSF and reduces the need
  for frame selection, thus leading to higher Strehl and improved
  sensitivity. Image reconstruction {\it a posteriori}, the
  possibility to use multiple reference stars and the fact that these
  reference stars can be rather faint means that holographic imaging
  offers a simple way to image large, dense stellar fields near the
  diffraction limit of large telescopes, similar to, but much less
  technologically demanding than, the capabilities of a
  multi-conjugate adaptive optics system. The method can be used with
  a large range of already existing imaging instruments and can also
  be combined with AO imaging when the corrected PSF is unstable.
 \end{abstract}

\begin{keywords}
instrumentation: high angular resolution, instrumentation: adaptive
optics, atmospheric effects, methods: observational, techniques: high
angular resolution, techniques: image processing
\end{keywords}

\section{Introduction}

Obtaining images at or near the diffraction limit of 4-10m class
telescopes from the ground is key to many astrophysical
projects. Before the end of the 1990s, this goal was generally reached
by speckle interferometry, also called speckle imaging, a technique
based on recording long series of short exposure images and digital
image reconstruction {\it a posteriori} with a range of different
algorithms
\citep[e.g.,][]{Labeyrie:1970fk,Knox:1976vn,Weigelt:1977uq,Lohmann:1983kx,Christou:1987fk,Christou:1991kx}.

\begin{table*}
\caption{Summary of observations used in this work.} 
\label{Tab:Obs}
\begin{tabular}{@{}lllcccccc} 
\hline 
ID & Target & Instrument & UTC Date & $\lambda_{c}$\,[$\mu$m]   &  $DIT$\,[s] &
$N_{\rm Frames}$ & Seeing\,[$"$] & Strehl\\
\hline
1 & Galactic centre & NaCo/VLT & 7 Aug 2011 & 2.18 & 0.15 & $12500$ & $0.6-0.9$ & $0.82$\\
2 & Galactic centre & NaCo/VLT & 31 Mar 2009 & 2.18 & 1.0 & $1920$ & $0.4-0.5$ & $0.34$\\
3 & NGC\,3603 & NaCo/VLT & 29 Jan 2010 & 1.27 & 0.11 & $4928$ & $0.5$ & $0.4$ \\
4 & NGC\,3603 & NaCo/VLT & 03 Aug 2010 & 2.18 & 0.4 & $4928$ & $0.6$ & $0.13$ \\
5 & M15 & FASTCAM/NOT & 23 Oct 2008 & 0.872 & 0.03 & $50000$ & $1.0$ & $0.18$ \\
6 &Galactic centre & VISIR/VLT & 22/23 May 2007 & 8.59 &  0.02 &
$\sim$$100000$ & $2.0-3.0$ & $>0.9$ \\
7 & 47\,Tuc & NaCo/VLT & 1 Aug 2010 & 2.18 & $3.0$ & $1900$ & $1.0$ & $>0.9$ \\
8 & M\,30 & HAWKI/VLT & 30 Apr 2012 & 2.15 & $0.2$ & $128$ & $0.9$ &  \\
\hline
\end{tabular}
\medskip

$\lambda_{c}$ is the central wavelength of the broad-band filter
used. $DIT$ (detector integration time) is the
exposure time of  individual frames. $N$ is the total number of
frames obtained.  The {\it Strehl} column refers to the approximate Strehl
ratio (uncertainty $\sim$$0.05$) reached
in the optimal image reconstructed via our holography algorithm or  in
the final AO image  (only for ID\,2). The column titled {\it
  Seeing} lists the approximate visual seeing during the
observations. No Strehl estimate is given for the observations of
M\,30 because the diffraction limit was strongly under-sampled in the
observations (see section\,\ref{sec:under}). Data sets ID\,2 and ID\,7
were obtained with, the rest without AO. 
\end{table*}

During the past two decades adaptive optics (AO) assisted observations
have become the standard for obtaining near-infrared (NIR) images near the
diffraction limit of large telescopes. An important advantage of AO is that the real-time image correction
makes long integration times possible. This increases significantly
the sensitivity of the observations, which is limited by the detector
readout noise in speckle imaging, and also allows one to use
techniques like coronagraphy or spectroscopy.

However, in spite of the enormous success of AO instrumentation, the
technique is not perfect. A broad range of systematic effects 
can lead to difficulties in calibrating the AO PSF \citep[see the
ample discussion in][]{Tuthill:2006fk}. Another difficulty is that the
PSF of an AO image varies systematically across the field-of-view
(FOV) and that the Strehl ratio deteriorates rapidly with increasing
distance from the guide star. These {\it anisoplanatic} effects result
because different viewing directions probe different turbulence
profiles of the atmosphere, but a single-conjugated AO system only
corrects the wavefront toward the guide star.

Various solutions to these problems have been developed, like
AO-assisted sparse aperture masking
\citep[SAM,][]{Tuthill:2006fk,Lacour:2011fk} or multi-conjugate
adaptive optics \citep[e.g.,][]{Rigaut:2000fk}. In addition, the use
of laser guide stars (LGSs) frees AO systems from the requirement of
having bright stars close to the target. Nevertheless, all these
solutions come at a price. Greater technical complexity is inevitably
accompanied by rising costs and increased vulnerability of the
systems. It is therefore important to keep exploring the possibilities
of speckle imaging, which has, for example, recently experienced a
renaissance at optical wavelengths, where AO still cannot work
reliably, in the form of {\it lucky imaging} (see
section\,\ref{sec:methods}).

In this paper we describe a new implementation of the so-called {\it
  speckle holography} technique.  We have written a program package
that provides excellent results based on iteratively improved
extraction of the instantaneous PSF from speckle frames and on the
optional simultaneous use of multiple reference stars. The algorithm
has been specifically developed for crowded fields, but work also on
isolated targets if at least one reference star is located
sufficiently close. The algorithm shows excellent performance from the
NIR to optical wavelengths, works with faint guide stars, and is more
efficient than lucky imaging techniques. We tested our method on a
broad range of targets, with different telescopes and instruments, and
under various conditions. Table\,\ref{Tab:Obs} summarizes the
observations used in this work.\footnote{Based on observations made
  with ESO Telescopes at the La Silla Paranal Observatory under
  programmes 060.A-9800(J,L), 087.B-0658(A), 179.B-0261(X), and
  485-L.0122(A).} All data were reduced in a standard way (sky
subtraction, dead pixel correction, flat-fielding).

\section{Speckle image reconstruction} \label{sec:methods}

\begin{figure}
  \caption{\label{Fig:reduction} 
    $Ks$ speckle observations of the Galactic centre. Top: Single speckle
    frame with reference stars used for reconstruction marked by
    circles (due to dithering one stars lies near the edge of the
    field shown here). Middle: SSA image. Bottom: Holographic
    reconstruction. Insets: PSFs for each image.  All color scales are
    logarithmic.}
\includegraphics[width=.9\columnwidth,angle=0]{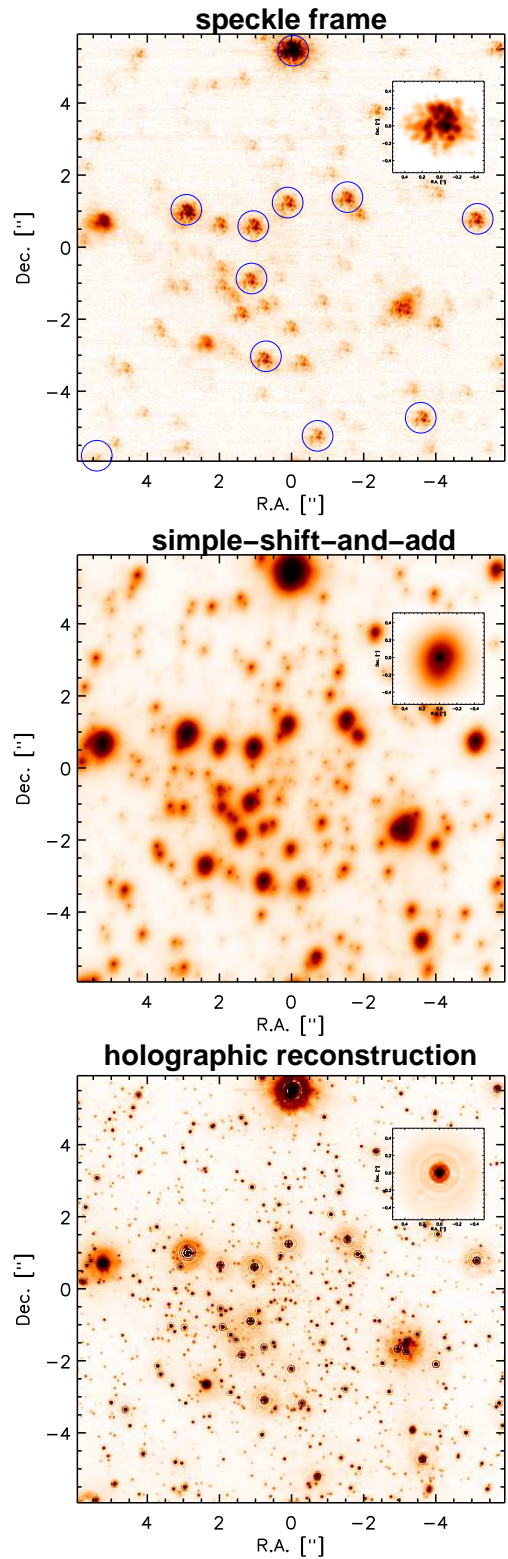}
\end{figure}

A range of different algorithms exists for image reconstruction from
speckle data. Among those, the so-called simple shift-and-add (SSA) method became
one of the most widespread methods
\citep[e.g.,][]{Christou:1991kx,Eckart:1994fk}.
In a simplified view, each speckle in the instantaneous PSF, or {\it
  speckle cloud} (see inset in top panel of
Fig.\,\ref{Fig:reduction}), can be regarded as a diffraction-limited
image of the source. The SSA method consists of applying a shift,
given by the xy-offset of the brightest pixel of a reference star's
speckle cloud from a chosen reference pixel, to each sequential image
in the stack before averaging the frames.
The PSF in the resulting image can then be described by the
superposition of an Airy function over a broad Gaussian seeing halo.
With this method, typical Strehl ratios of the order 10\% can be
achieved at $\lambda=2.2\,\mu$m at an 8-m telescope. 

The SSA algorithm became very popular because it 
is easy to understand, fast, and robust. On the downside, SSA makes inefficient use of the information and
photons contained in the individual exposures. Thus the majority of
the photons of a source end up in its seeing halo, which leads to
low-Strehl, low-sensitivity images.

When combined with rigorous selection of the best frames,
significantly higher Strehl ratios can be reached. This is due to the
statistical nature of turbulence, which has the effect that a small
percentage of the speckle frames will resemble diffraction limited
short exposures, with little distortion by the atmosphere. This method
is termed {\it lucky imaging} and has recently become popular at
optical wavelengths
\citep[e.g.][]{Hormuth:2008fk,Labadie:2010fk}. While working well at
small telescopes, lucky imaging becomes increasingly inefficient with
larger telescope apertures because the mean number of speckles
increases quadratically with telescope diameter. Thus, it can become
necessary to discard $90-99\%$ of the data in order to achieve a
diffraction-limited image in the optical regime.

A more efficient image reconstruction algorithm is the so-called {\it
  speckle holography} technique. The algorithm is based on a division of averaged quantities in Fourier
space:
\begin{equation}
\label{equ:holo}
O = \frac{\langle I_{m}P^{*}_{m}\rangle}{\langle\vert P_{m}^{2}\vert\rangle},
\end{equation}
where $O$ is the Fourier transform of the object, $I_{m}$ and $P_{m}$
are the Fourier transforms of the $m$-th image (speckle frame) and of
its instantaneous PSF, respectively, and the brackets denote the mean
over $N$ frames.  $P_{m}^{*}$ is the conjugate complex of $P_{m}$. Taking
the average over a large number of frames will effectively suppress
the noise.

It can be shown that equation\,\ref{equ:holo} is the best estimate, in
the least squares sense, of the Fourier transform of the object
\citep{Primot:1990fk}. For an infinite number of frames, the
denominator becomes equal to the speckle transfer function related to
the turbulence. This feature of the algorithm allows one to
reconstruct a high-fidelity image from the combination of typically
several hundreds to thousands of speckle frames. The actual image is
then obtained after apodising $O$ with the optical transfer function
of the telescope (OTF, usually an Airy function) to suppress power
above the telescope cut-off frequency and applying an inverse Fourier
transform \citep[see][for a flow-diagram of the
algorithm]{Petr:1998vn}. The great advantage of speckle holography
over the SSA technique is that it uses the {\it full} flux and
information content of each speckle cloud. It thus results in
significantly reduced seeing halos around point-sources and in higher
Strehl ratios than the SSA technique.  Frame selection is generally
not necessary, which leads to a much higher observing efficiency
($\sim100\%$ of frames used) than in the case of lucky imaging
($\sim10\%$ of frames used, depending on seeing and telescope size).
This increases significantly the sensitivity and dynamic range of the
reconstructed images.

Key to a successful application of holography is the reliable
extraction of the instantaneous PSF from each speckle frame. This
poses no problem if a bright star is located close to the target,
i.e.\ within an isoplanatic angle, and if this star is isolated, i.e.\
if there are no secondary sources present within the area covered by
the speckle clouds. However, this ideal situation is rare.
Secondary sources close to the reference star will lead to biases in
the extracted instantaneous PSFs, which will result in systematic
errors, such as the presence of positive or negative spurious sources
(``ghosts''), in the reconstructed image.  This poses an important
obstacle to the use of speckle holography, particularly in crowded
fields. Therefore, holography was in practice rarely
used. Alternatively, {\it bispectrum}, or speckle masking, has been
more commonly used because this technique allows one to reconstruct an
image of complex targets without the necessity of observing an
unresolved reference source
\citep[e.g.,][]{Weigelt:1977uq,Lohmann:1983kx,Weigelt:2006fk}. But
bispectrum methods are complex and require considerable amounts of
computing time.

Here, we present an improved algorithm that addresses the key problem
of speckle holography -- the extraction of reliable
instantaneous PSFs -- in a twofold way: (a) the optional use of several
reference stars {\it simultaneously}, which suppresses systematic
errors caused by secondary sources near the reference stars, and (b)
an {\it iterative} approach to PSF extraction, that uses the relative
positions and fluxes of sources known from previous image
reconstructions for the next iteration.

\section{Methodology} \label{sec:algorithm}

Our method is illustrated using the example of the reduction of
speckle data of the Galactic centre (GC, ID\,1 in
Table\,\ref{Tab:Obs}).  The data were acquired with NaCo/VLT in the
so-called {\it cube mode} \citep[see][]{Girard:2010fk}, that allows
for the fast recording of a series of short exposures with a minimal
overhead. Each cube comprised 500 frames.  Visual seeing varied
between $\sim$$0.6"-0.9"$ and the atmospheric coherence time was
$\tau_{0}\approx2\,$ms, i.e.\ seeing was fast. We proceeded as
follows:
\begin{enumerate}
\item Creation of long exposure images for each data cube. 
\item Pre-alignment of all cubes to remove dithering offsets.
\item Image reconstruction with the SSA algorithm (middle panel of
  Fig.\,\ref{Fig:reduction}). Fine-alignment of the frames by using the
  centroids of the speckle clouds of the SSA reference star. 
\item  Astrometry and photometry with {\it StarFinder}
  \citep{Diolaiti:2000fk} on the SSA-image. This provides relative
  positions (to sub-pixel accuracy) and fluxes of the stars which are
  needed to estimate (see vi) and iteratively improve the PSFs (see
  viii) for each frame. 
\item Selection of reference stars (see top  panel of
  Fig.\,\ref{Fig:reduction}). 
\item Estimation of the
  instantaneous PSF for each speckle frame
  from the median of the sub-pixel aligned, flux-normalized
  images of the reference stars.
\item Noise thresholding: Estimation of noise, $\sigma$, and constant
  background, $bg$, for each PSF from its mean and standard deviation
   in a sufficiently large annulus around its centre.  A
  constant value of $bg+n\times\sigma$ (typically, $n=1-3$) is
  subtracted from the PSF. All pixels that acquire negative values
  through this subtraction are set to $0$.  Finally, a circular mask
  is applied to the PSF and it is normalized to a total flux of one.
\item Subtraction of all known secondary, contaminating sources near
  the reference stars in each frame, using the preliminary
  PSFs and information from step (iv). Subsequently, improved PSF
  estimates are obtained
  for each frame (see inset in Fig.\,\ref{Fig:reduction}) by
  applying again steps (vi)
  and (vii). 
\item $O$ is estimated by applying equation\,\ref{equ:holo}.
\item Apodization with the $OTF$. The $OTF$ was assumed to be an Airy
  function, and was constructed with the routine {\it strehl}
  \citep[ESO software package {\it eclipse}, ][]{Devillard:1997kx}.
\item Inverse Fourier transform to obtain the reconstructed
  image (bottom panel of Fig.\,\ref{Fig:reduction}). 
\item Repetition of the process, starting at step (iv) but using the
  holographically reconstructed image, which is of significantly
  higher quality than the initial SSA image. This means more stars are
  detected, with greater accuracy of their positions and
  fluxes. Reference stars that appear as close multiples or elongated
  in the holography image can be discarded. The result of this last
  iteration is shown in the bottom panel of Fig.\,\ref{Fig:reduction}.
 \end{enumerate}

 Although the mean in equation\,\ref{equ:holo} is very efficient in
 suppressing random noise, systematic errors can be present in the raw
 PSFs that are obtained from the median superposition of the reference
 stars. Sources of such systematic errors can be, e.g., the finite
 accuracy with which contaminating sources can be subtracted or
 features produced by the detector or readout electronics. Any
 additive offset to the PSF that may be caused by smooth, extended
 emission in the target must be taken into account. For these reasons
 it is important to apply the background subtraction and noise
 thresholding in step (vii). Mainly due to this noise thresholding,
 some of the flux in the reconstructed image ends up in halos around
 the stars. The noise thresholding would be unnecessary if the
 reference star(s) were perfectly isolated and there were no sources
 of systematic uncertainties.

 The SSA image (Fig.\,\ref{Fig:reduction}, centre) has an
 estimated Strehl ratio of $\sim$$9\%$, while the
 holographic reconstruction (Fig.\,\ref{Fig:reduction}, bottom)
 leads to $\sim$$82\pm5\%$,. The Strehl and its uncertainty were
 estimated from
 three  PSFs created with {\it StarFinder} using three
 different, non-overlapping sets of reference stars.

\subsection{Quality control}

Details of our astrometric and photometric examination of the image
resulting from our holographic reconstruction method are given in
appendix\,\ref{sec:uncert}. Here we only note that the image has an
extremely high Strehl ratio and very good PSF cosmetics. Photometric
uncertainties are $\leq1\%$ and astrometric uncertainties
$\leq0.01$\,pixel for stars having $Ks\leq15$. Spatial and temporal
PSF stability are high. Simulations show no systematic bias in the
astrometry and photometry of point sources. Moreover, the noise
properties of the image are largely Gaussian and no significant
cross-correlations between the pixels could be detected.

\subsection{Comparison with AO imaging \label{sec:AOcomp}}

\begin{figure}
\includegraphics[width=\columnwidth,angle=0]{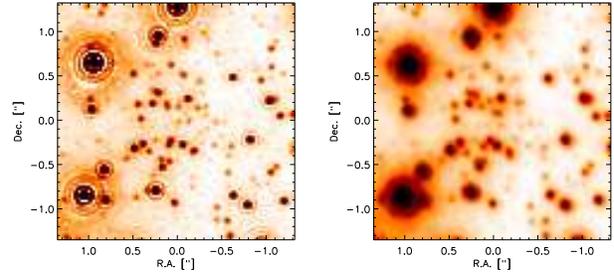}
\caption{\label{Fig:holoAOim} Zoom onto a $\sim2.5"\times2.5"$ region
  centred on Sgr\,A* (NaCo/VLT, $Ks$). Left: Image reconstructed from
  speckle data (ID\,1 in Table \,\ref{Tab:Obs}) with our holography
  algorithm. Right: AO assisted image (ID\,2 in Table
  \,\ref{Tab:Obs}). Note that some of the fast-moving stars near
  Sgr\,A* are located at different positions in the two images, which
  were taken more than 2 years apart.}
\end{figure}

\begin{figure}
\includegraphics[width=\columnwidth,angle=0]{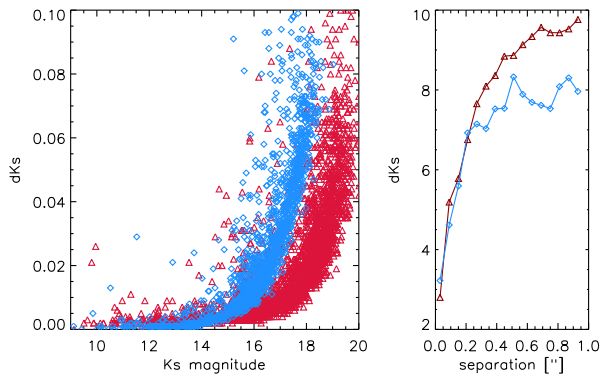}
\caption{\label{Fig:GCHoloAO} Photometry in the holographic (blue
  diamonds) and AO (red triangles) $Ks-$band images of the GC. Left:
  Photometric uncertainty vs. magnitude. Right: Dynamic range
  vs. separation for stellar pairs. }
\end{figure}

We compared the holographic image of the GC with a $Ks-$band AO image
of the same region. The quality of the AO data chosen here (ID\,2 in
Table \,\ref{Tab:Obs}) is exceptionally high, with visual seeing of
only $0.4"-0.5"$ and an extremely long coherence time of $\tau_{0}\sim$47\,ms.
The loop of the AO was closed on IRS\,7, the brightest star visible in
the image shown in the panel (a) of Fig.\,\ref{Fig:reduction}.  A comparison
between the holographically reconstructed image from August 2011 and
the AO image from March 2009 is shown in Fig.\,\ref{Fig:holoAOim},
which shows a  zoom onto the region surrounding Sagittarius\,A* (Sgr\,A*).

A quantitative comparison of PSF fitting photometry in the AO and in
the holography images is shown in Fig.\,\ref{Fig:GCHoloAO}. The
analysis was done on a $13"\times13"$ field centred on Sgr\,A*. The
detection threshold of {\it StarFinder} was set to $5\,\sigma$ and the
correlation threshold to $0.9$ in order to suppress the detection of
spurious sources. The {\it noise map} method, as described in
appendix\,\ref{sec:uncert} was used to estimate the pixel
uncertainties in both images. As can be seen in the left panel of
Fig.\,\ref{Fig:GCHoloAO}, the AO image is about $1.5$\,mag deeper than
the holography image. Note that this is hard to see in
Fig.\,\ref{Fig:GCHoloAO}, which shows the most crowded region of the
field, around Sgr\,A*, where the completeness limit
is $Ks\sim18$.  It is interesting to note that the photometric
uncertainty for stars $Ks\leq14$ in the holography image is smaller
than in the AO image.  The right panel of Fig.\,\ref{Fig:GCHoloAO}
shows the dynamic range for the detection of close point sources in
both images.  The figure shows the median of the five stellar pairs
with the highest dynamic range in each bin of $0.06"$ width (one
resolution element). At separations $<0.2"$, the dynamic range of the
holography image is equivalent to that of the AO image. We believe
that the larger dynamic range of the AO image at separations $>0.2"$
is mainly caused by the higher sensitivity of the AO image, which
contains a far greater number of faint stars.

For an exact quantitative comparison between holography and AO
the data should be taken under the same atmospheric conditions with
the same on-source integration times, of course. We note, however,
that the comparison shown here is a conservative one, favouring the AO
image, because seeing was significantly better for the AO
observations. In particular, the coherence time was a factor $>10$
longer than for the speckle observations and unusually long for
conditions at Paranal ($\tau_{0} \leq 6.6$\,ms $95\%$ of the time in
the interval 1999-2003, as found on the ESO web site). We also note
that the magnitude of the guide star, IRS\,7, an irregular variable,
was $Ks=6.96\pm0.04$ (measured after repairing its saturated core with
{\it StarFinder}) during the AO observations,  $\sim$$0.5$\,mag
brighter than at other times \citep[$Ks=7.69\pm0.06$ in April
2006,][]{Schodel:2010fk}, which will have resulted in improved AO
performance compared to other epochs. 

A further comparison between the performance of SSA, AO, and
holographic imaging is shown in appendix\,\ref{sec:Keck}, where we
describe the application of holographic imaging to NIRC/Keck GC
imaging data. In fact, it was NIRC/Keck data on which we developed and
fine-tuned large parts of our software.

\section{Holography under extreme conditions}

In this section we explore the limits of holographic imaging under
different, extreme conditions, to demonstrate its great utility with a
broad range of instruments and under a great variety of scenarios. We
will also discuss questions such as the capability of holography to
reconstruct extended targets or to deal with anisoplanatic effects.

\subsection{Faint PSF reference stars  \label{sec:faint}}

The faintest stars barely visible on the individual speckle frames
taken on the GC have $Ks\approx13$. We successfully tested image
reconstruction by using a single, isolated star of $Ks=12$ and
obtained an image with very good image cosmetics and a Strehl ratio of
$\sim18\%$.  The limiting brightness for a natural guide star for
NaCo's AO is heavily dependent on atmospheric conditions. It is
roughly $Ks\approx12$, leading to an estimated Strehl $<10\%$. In a
further test, by combining 24 reference stars of $Ks=13\,\pm0.5$,
distributed throughout the field, a Strehl $\sim$$45\%$ was reached
(Fig.\,\ref{Fig:tests}, left). This ability to work with (multiple)
faint, NIR reference sources supersedes the possibilities of most, if
not all, currently existing AO systems.

\subsection{Centre of correction}

\begin{figure}
\includegraphics[width=\columnwidth,angle=0]{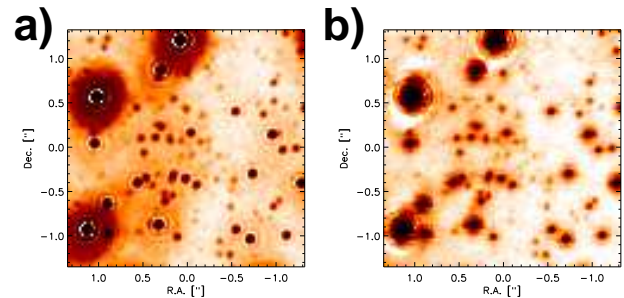}
\caption{\label{Fig:tests} (a) Holographic reconstruction of the
  NaCo/VLT $Ks$-band GC speckle data with   24 reference stars of
  $Ks\approx13$, distributed throughout the field, within about $15''$
  of Sgr\,A*, which is located at the origin of this image. (b)  Holographic
  reconstruction of the same image, but using only the bright
  supergiant IRS\,7, located $\sim$$5.5"$ north of Sgr\,A*.}
\end{figure}

Contrary to AO with a single,  natural guide star, holographic imaging of a
dense cluster allows the observer to choose the centre of optimal
correction. As an illustration, we reconstructed the image using only
the brightest star in the field, IRS\,7, as reference source. Since
IRS\,7 is located approximately $5.5"$ north of Sgr\,A*, the
correction is not optimal in the environment of the black hole
(Fig.\,\ref{Fig:tests}, panel (b)), in contrast to the images shown in
Fig.\,\ref{Fig:holoAOim}, panel (a) or in Fig.\,\ref{Fig:reduction},
where the reference stars were chosen distributed throughout the
field around Sgr\,A*.

\subsection{Extended sources}

Reconstruction of extended sources presents no difficulties. This has
been demonstrated by \citet{Schodel:2011uq}, who created
holographically reconstructed high-Strehl mid-infrared (MIR) images of the
interstellar medium at the GC.

\subsection{Long exposure times \label{sec:long}}
 
Since the short exposure times necessary for speckle imaging require
windowing on some detectors (e.g. $512\times512$ on the NaCo detector
for $DIT=0.1$\,s), a question of high interest is whether the speckle
imaging technique can also work with exposure times significantly
longer than an atmospheric coherence time. To test this, we observed
NGC\,3603, using $DIT=0.4$\,s (ID 4 in Table \,\ref{Tab:Obs}),
which allowed us to take advantage of the full NaCo detector FOV. The
diffraction limit was reached with a Strehl of $\sim13\%$. We conclude
that diffraction-limited imaging is possible even with exposure times that exceed the atmospheric
coherence time by factors $\gg10$ ($\tau_{0}$ was a few ms
during the observations), albeit at the cost of a lower Strehl in the
final image.

\subsection{Short NIR and optical wavelengths \label{sec:optical}}

The reconstructed image from $J-$band speckle  observations of
the core of NGC\,3603 (ID 3 in Table \,\ref{Tab:Obs}) is shown in the
left panel of Fig.\,\ref{Fig:short}. With the three brightest stars
(magnitudes $J\approx9$) used as reference sources, a Strehl of
$\sim$$40\%$ was achieved. This supersedes the predicted performance
of NaCo's AO. The {\it NaCo preparation software
  (http://www.eso.org/sci/observing/phase2/SMGuidelines/-NAOSPS.html)}
can be used to get a rough idea of AO performance in the $J-$band. It
predicts an on-axis Strehl ratio of about $20\%$ under similar
atmospheric conditions and with a {\it visual} guide star of $V=9$.
With an infrared guide star of $J=9$ the predicted on-axis Strehl is
 about $13\%$.

 In order to test the performance of holography in the optical regime,
 we used $I-$band observations of the core of the globular cluster
 M15, obtained with FASTCAM \citep{Labadie:2010fk} at the 2.5-m Nordic
 Optical Telescope (NOT) (ID 5 in Table\,\ref{Tab:Obs}).  After
   selecting the best $\sim$$1\%$ of the data, a Strehl of $\sim$$4\%$
   and a dynamic range of about $5$\,magnitudes (assuming a
   $5\,\sigma$ detection limit in {\it StarFinder}) were achieved with
   the {\it lucky imaging} technique. With holography the result on
   the same set of selected frames could be considerably improved,
   reaching a a Strehl of $\sim$$27\%$ with a dynamic range of about
   $6.3$\,magnitudes. Selection of the speckle frames with the best
   PSFs is not necessary for holography. Therefore, by using the best
   $50\%$ of the frames, we were able to increase the dynamic range of the
   holographic image even further, to about $8.0$\,magnitudes. Using an even
   larger amount of frames was not possible for the data used here
   because they were taken under very bad and highly variable seeing
   conditions, so that even the brightest stars were barely visible in
   many frames.

\subsection{Anisoplanatic effects \label{sec:aniso}}

\begin{figure}
\includegraphics[width=\columnwidth,angle=0]{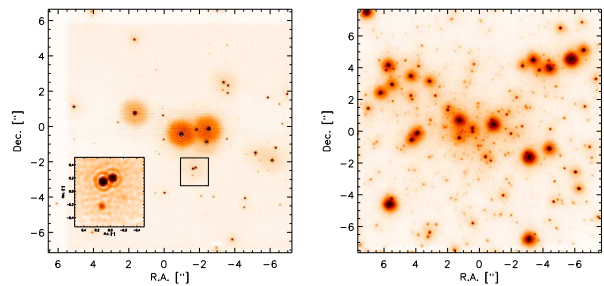}
\caption{\label{Fig:short} Left: Reconstructed J-band image of the
  core of NGC\,3603. The inset shows a zoom of the region marked by
  the black box. The two close stars in the box are separated by
  $0.078"$ and have $J=12.0$ and $J=12.6$. Right: Reconstructed I-band
  image of the core of M15. Five overlapping subfields were
  reconstructed separately to counteract anisoplanatic effects.}
\end{figure}

In the $I-$band, the isoplanatic angle is only on the order of a few
arcseconds, while the FOV of FASTCAM is about $15"$. We divided the
FOV into five overlapping subfields, each of size about $9.5"$.
Holographic reconstruction was applied to these subfields separately,
using several reference stars in each one. The final image was created
by mosaicking the reconstructed subfields together. The optimal shifts
between them were determined via cross-correlation.  The corresponding
image has a highly homogeneous PSF across the field and is shown in
the right panel of Fig.\,\ref{Fig:short}. About 10 stars with
$I=12-13$, distributed over the field, were used as reference sources
for the holographic reconstruction.

\subsection{MIR wavelengths \label{sec:MIR}}

Speckle techniques are naturally suited to improve images in the MIR
because of the short readout times necessary at these wavelengths.  In
\citet{Schodel:2011uq} we reconstruct fully diffraction limited
(FWHM$\approx0.25"$), high-Strehl ($\geq90\%$) images of the Galactic
centre from VISIR/VLT $8.6\,\mu$m speckle imaging data during visual
seeing as bad as $\sim$$2-3"$. This shows that in the MIR, 
holography allows one to obtain high-Strehl images even under the most
adverse seeing conditions. This can greatly enhance the flexibility
and efficiency of such observations at large telescopes.

\subsection{Reference sources with extended
  emission \label{sec:extref}}

Holography can also work if the reference source is surrounded by
diffuse emission.  The reference star used for the MIR imaging
described above, IRS\,3, sits atop extended, diffuse emission. In that
case it is merely necessary to set the noise threshold in the PSF
determination high enough to suppress the contribution of the diffuse
emission \citep[see][for more details]{Schodel:2011uq}.  Increasing
the noise threshold will make holography, which uses the entire PSFs,
ever more similar to the {\it weighted shift-and-add} method, which
uses only the relatively weighted local maxima of the instantaneous
PSFs \citep{Christou:1987fk}.

\subsection{Holography with AO \label{sec:AO}}

\begin{figure}
\includegraphics[width=\columnwidth,angle=0]{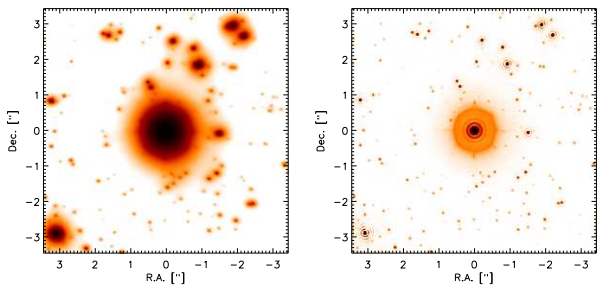}
\caption{\label{Fig:AOplusholo} Left: AO-assisted  $Ks-$band image of a
  field centred on the star 2MASS 00241142-7205556 in the globular
  cluster 47\,Tuc. Right: Holographic image reconstruction of the same data.}
\end{figure}
 
The necessarily short exposure time in speckle imaging limits the
sensitivity to faint sources. Longer exposure times become possible
with the help of AO. However, application of the holography technique
will only be preferable if the AO PSF is {\it not completely} stable,
i.e.\ if the PSF shows some random variability on time scales similar
to the exposure time. In that case, the holography technique can
efficiently suppress the random fluctuations and lead to high-Strehl
images.

We have tested this {\it holography plus AO} technique on imaging data
of a field in the globular cluster 47\,Tuc (ID 7 in
Table\,\ref{Tab:Obs}). In these data the PSF is strongly variable
between the frames. In Figure\,\ref{Fig:AOplusholo} we show a
comparison between a standard reduction and a holographic image
reconstruction of the same set of frames. The central star in the
field was used as reference source. Application of holographic
reconstruction improves the resolution of the image, without the
strong ringing of Wiener deconvolution or the grainy background of
Lucy-Richardson deconvolution. Hence, just as the performance of SAM
can be boosted by the stabilizing effects of AO, so too can speckle
holography. Holography can be very useful when AO correction is unstable, for
example, under bad seeing conditions or at short wavelengths.

\subsection{Under-sampling: larger FOV and improved sensitivity \label{sec:under}}

\begin{figure}
\includegraphics[width=\columnwidth,angle=0]{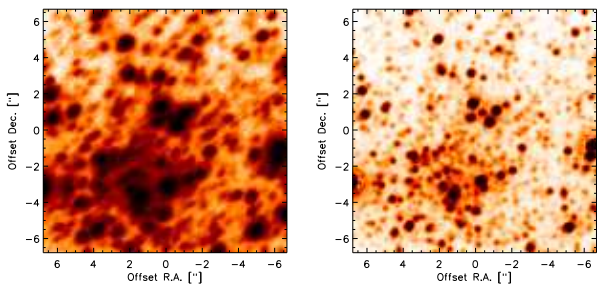}
\caption{\label{Fig:HAWKI} $Ks$-band imaging of M30 with
  VLT/HAWKI. Left: Standard long exposure with a resolution of
  $\sim$$0.7"$. Right; Holographic reconstruction with a resolution
  of $\sim$$0.27"$.}
\end{figure}

Using holography on imaging data that under-sample the diffraction
limit can have two important advantages. First, each detector pixel
will receive more photons, thus improving the sensitivity of speckle
observations. Second, a much larger instrumental FOV can be
achieved. The angular resolution that can in this case be reached is,
of course, limited by the spatial frequency sampled adequately by the
pixel scale. But this can still be significantly better than the limit
imposed by atmospheric seeing. We tested this idea with data acquired
by the HAWKI NIR widefield camera installed at the ESO VLT on the
globular cluster M30 (ID\,8 in Table\,\ref{Tab:Obs}). An image of the
cluster centre that results from simple, incoherent averaging of all
exposures (corresponding to a standard long-exposure) is shown
in the left panel of Figure\,\ref{Fig:HAWKI}.  It has a PSF FWHM of
roughly $0.7"$. Since the diffraction limit was under-sampled, we
restored the holographic reconstruction with a Gaussian PSF of
$2.5$\,pixel FWHM (corresponding to $0.27"$ with the HAWKI pixel
scale). The holographic reconstruction is shown in the right panel of
Figure\,\ref{Fig:HAWKI}. The image sharpness was improved by more than
a factor of two. Stars as faint as $Ks\sim20$ could be detected in
spite of the short exposure time ($0.2$\,s) and total integration
time ($26$\,s).

For technical reasons (non-optimal setup and very short observations,
large dithering), we only show a  small field in
Fig.\,\ref{Fig:HAWKI}. Note, however, that with $DIT=0.2$\,s the HAWKI
detector can be read out in a $512\times2048$ pixel field,
corresponding to a FOV of $217"\times54"$ or $3.25$ square
arc-minutes. Moreover, stars as faint as $Ks\sim16$ can be detected in
the speckle frames and thus serve as reference for image
reconstruction. Since holography  can work with
significantly longer exposure times (see section\,\ref{sec:long}),
even larger FOVs and fainter guide stars will be feasible. 

It may be possible to even recover some of the information lost by
under-sampling of the diffraction limit. When the condition of Nyquist
sampling is not fulfilled, the detailed appearance of the speckle
clouds of different stars in the field will depend significantly on
the relative pixel positions of the stars. However, using a large
number of exposures and multiple reference stars, one may recover some
of the lost information, similar to dithering by half-pixels in
observations with the Hubble Space Telescope's wide-field NIC3 camera
\citep[e.g.][]{Wang:2010fk}. However, we have not yet tested this
possibility.

\section{Summary and Discussion}

An optimized method for applying holographic reconstruction to speckle
imaging data of dense stellar fields has been presented. The two key
features of our methodology are an iterative improvement of the
instantaneous PSFs extracted from each speckle frame and the
(optional) use of several reference stars simultaneously. The method
has been tested with great success on a range of different
instruments, wavelengths and targets.  Excellent PSF cosmetics and
high Strehl ratios have been achieved. The results show that
holographic imaging can in many cases compete with AO imaging and even
outperform it, particularly at short wavelengths. In the following, we
briefly discuss the advantages and disadvantages of speckle holography
in given observing situations, particularly with respect to
alternatives, like AO-assisted imaging or SAM.

\subsection{Sensitivity}

The sensitivity in holographic imaging is usually limited by detector
readout noise. Note that, to zeroth order, sensitivity is independent
of the telescope diameter, $D$, because both the number of photons
collected and the number of pixels needed to sample the diffraction
limit increase as $D^{2}$. 

However, the constraints on sensitivity are not severe: In just
$1900$\,s of total integration time, we have reached a $3\,\sigma$
point-source sensitivity of $Ks\approx19$ in observations of the GC
with the S27 camera of NaCo/VLT. The sensitivity of holographic
imaging is clearly superior to the SSA/lucky imaging method. It is far
more sensitive than SAM, even if the latter is
supported by AO, because the masks block on the order of $90\%$ of the
light. The sensitivity of holography means that the technique can be
well suited to study star clusters in the Milky Way. Main sequence
stars of G-type can be detected out to 10\,kpc; pre-main sequence
stars at the H-burning limit with an age $<10^7$\,yr can be detected
out to $\sim$3\,kpc.

Sensitivity can be improved by optimizing readout electronics for
short exposure times. For example, the group of G.\,Weigelt (MPIfR,
Bonn, Germany, priv. comm.) has reached a readout noise on the order
of $10e^{-}$ in their speckle camera, using a HAWAII-1 detector. NaCo
uses the same detector type, but has a readout noise that is almost 4
times higher (in the fast readout modes required for speckle
imaging), corresponding to a reduction of sensitivity of almost 1
magnitude. The next generation of fast, noiseless NIR detectors will
provide a sensitivity on the order of 3 magnitudes deeper
\citep[e.g.][]{Finger:2010uq,Figer:2011fk}.

As we show in section\,\ref{sec:under}, under-sampling of the
diffraction limit is another, rather easy, way to improve
sensitivity. Under-sampling the diffraction limit by a factor of
$\sim$3 (corresponding to a resolution $\sim$$0.18"$ at an 8m-class or
$\sim$$0.05"$ at a 30m-class telescope in the K-band) would increase
sensitivity by more than one magnitude.

Finally, for a given sky brightness, the number of sky photons per
pixel will depend on angular resolution. At small $D$, and therefore
large pixel scale, infrared images will be sky-limited at far shorter
readout times than with a large telescope. For example, with NOTCam at
the 2.5\,m NOT, noise is limited by the sky already for readout times
$\leq0.5$\,s, assuming a sky brightness of
$Ks=$13\,mag\,arcsec$^{-2}$. This means that the short readout times
required by holography will not lead to any significant loss in
sensitivity in the infrared at small telescopes.

\subsection{Reference stars}

Holography can achieve high Strehl ratios even with relatively faint
($Ks\approx13$) reference stars, particularly when multiple reference
stars can be combined. No special infrared wavefront sensor is
needed. Holography is thus naturally suited to peer into the most
obscured places of the Galaxy (e.g., its central region) which may even be
challenging for LGS observations because of a lack of tip-tilt stars.

\subsection{Anisoplanatic effects}

Image reconstruction {\it a posteriori} provides the great advantage
that anisoplanatic effects can be compensated, as we have demonstrated
in section\,\ref{sec:aniso}. All that is needed is a sufficient
density of reference stars in all parts of the field.  The FOV is
then, in principle, only limited by the detector size and the ability
to read the latter with sufficient speed.  In the case of AO, such a
feat would require extremely complex MCAO systems with multiple lasers
\citep[e.g.,][]{Neichel:2010fk}.  Thus, holography is ideal for
imaging large fields in Galactic targets such as young (and embedded)
clusters, the bulge, the Galactic centre, or globular clusters. We
believe that the method can be refined, for example by using
interpolation or principal component decomposition to determine the
instantaneous PSF at any location in the field. This is, however,
beyond the scope of this work.

\subsection{PSF calibration}

Speckle holography results in a well-calibrated PSF with good
cosmetics. A necessary condition is, however, the presence of either
an isolated bright star near the science target or, alternatively,
that the science target be embedded in a star cluster.  For bright,
isolated science targets, AO-assisted SAM is the
preferable technique because it can guarantee a highly accurate PSF
calibration using off-target calibration sources. Also, the dynamic
range of AO assisted SAM for the detection of companion point
sources around bright ($Ks<12$) stars can be as high as $\Delta
m\approx6.5$ at separations $\lambda/D$ \citep{Lacour:2011fk}, where
$D$ is the telescope diameter and $\lambda$ the observing
wavelength. Hence, at the smallest angular scales and for bright
targets ($Ks\leq10-12$ with NaCo, depending on AO correction) aperture
masking surpasses the capabilities of holography demonstrated here.

Modern telescopes are highly adaptive and are conceived with many
control loops (active optics, adaptive optics, etc.) which can limit
the performance of techniques that require fast switching between
science target and calibrator (e.g., SAM). For such techniques, it is
very important that the telescope optical transfer function stay
constant during the entire observation. This is rather difficult to
guarantee with a VLT-like (thin mirror) telescope and will probably be
nearly impossible with the next generation of extremely large
telescopes. Because of the possibility of {\it self-calibration},
speckle holography may therefore remain an attractive technique even with
the next generation of telescopes and AO systems.

\subsection{Short wavelengths}

Our experiments show that, at short wavelengths, holography
out-competes most, if not all, currently existing AO systems
as well as the lucky imaging approach. Since holography greatly
reduces the need for frame selection, it can significantly boost the
efficiency of optical speckle observations. Holography may currently
represent the best approach for high angular resolution imaging at
short NIR to optical wavelengths at telescopes of the 10m-class and
even for the future extremely large telescopes.

\subsection{Specialized techniques}

Holography is a technique that is optimal for imaging, but is
difficult to combine with additional techniques that require long
integration times with a stable PSF, like spectroscopy or
coronagraphy. Although it is possible to combine speckle imaging with
spectroscopy \citep[see, e.g.,][]{Genzel:1997dp}, use of an AO system
appears clearly to be superior  because of the far greater
sensitivity. Nevertheless, since a speckle camera can be built with a
very small number of mirrors, holography appears to be a technique
well suited for polarimetry.

AO-assisted SAM \citep[see][]{Tuthill:2006fk} and holography can be
highly complementary techniques. A good example is the science case of
examining multiplicity in stellar clusters. Under optimal conditions
SAM can provide a dynamic range of $\sim$500, or about $6.7$
magnitudes, at angular separations $\theta=\lambda/D$, where $\lambda$
is the observing wavelength and $D$ is the telescope diameter
($\theta\approx0.03"$ at the VLT in the $K$-band). However, aperture
masking will only work on the brightest targets and on small
fields. Holography can fill in data for larger fields and fainter
stars, albeit at larger angular separations.

\subsection{Speckle holography and AO}

A very interesting perspective is to combine the virtues of both AO
and holographic image reconstruction. Such an approach can be highly
advantageous in situations where the AO system can only partially
stabilize the PSF, for example when it operates at short wavelengths
or under very unstable seeing conditions.  We have provided a
demonstration of the latter application in section\,\ref{sec:AO}. Of
course, speckle holography can also be an extremely useful backup
method when an AO system fails to close the loop or suffers temporary
technical problems.

\subsection{Costs and effectiveness}

The greatest strengths of a speckle imaging system are arguably costs,
robustness, and reliability. Only fast readout and data storage
capabilities combined with a low-noise detector and electronics are
necessary. No complex, expensive systems that require long development
and debugging times are needed. The simplicity of a speckle system
also means that it is highly reliable and will work under a broad
range of circumstances. Additionally, a speckle camera will be of low
weight and thus pose little strain upon any telescope infrastructure.

A particularly attractive option is that holographic imaging can be
used with a large range of existing instruments, e.g.: VISIR, NaCo,
and HAWK-I (VLT); FASTCAM (NOT); ASTRALUX (CAHA); ASTRALUX SUR (NTT),
INGRID (WHT), NOTCAM (NOT, with the new readout electronics scheduled
for the end of 2012), or NIRC2 (Keck). This is probably just a small
subset of instruments having the necessary capabilities. We generally
recommend short readout times be implemented on any optical or
infrared imaging system that significantly over-samples the angular
resolution set by atmospheric seeing.

\section*{Acknowledgments}
RS acknowledges support by the Ram\'on y Cajal programme, by grants
AYA2010-17631 and AYA2009-13036 of the Spanish Ministry of Economy and
Competition, and by grant P08-TIC-4075 of the Junta de
Andaluc\'ia. Support for this work was provided by NSF grant
AST0909218 and the Levine-Leichtman Family Foundation.  RS thanks
Sridhar Rengaswamy for providing his speckle data simulation tool and
support in using it and Guillaume Montagnier and the ESO staff for
taking the NACO and HAWK-I data presented in this work. Optical data
are from observations made with the Nordic Optical Telescope in the
Spanish Observatorio del Roque de los Muchachos. We thank the
anonymous referee for reviewing this paper.

\bibliography{/Users/rainer/Documents/BibDesk/BibGC}

\newpage
\appendix

\section{Photometry and astrometry on images reconstructed via
  holography}
\label{sec:uncert}

The primary use of holographic imaging may be high precision
astrometry and photometry. We therefore dedicate this appendix to a
more detailed analysis of some of the main sources of
uncertainties. Also, this appendix serves to show that holography does
not (or hardly) suffer from some of the typical problems of standard deconvolution
procedures, like for example cross-correlations between pixels,
ringing, or the creation of spurious sources. We introduce a new method
for estimating the uncertainties of pixel values in images that can be
of general interest, not just for holography.

\subsection{Systematic effects}
\label{sec:sys}

\begin{figure}
\includegraphics[width=\columnwidth,angle=0]{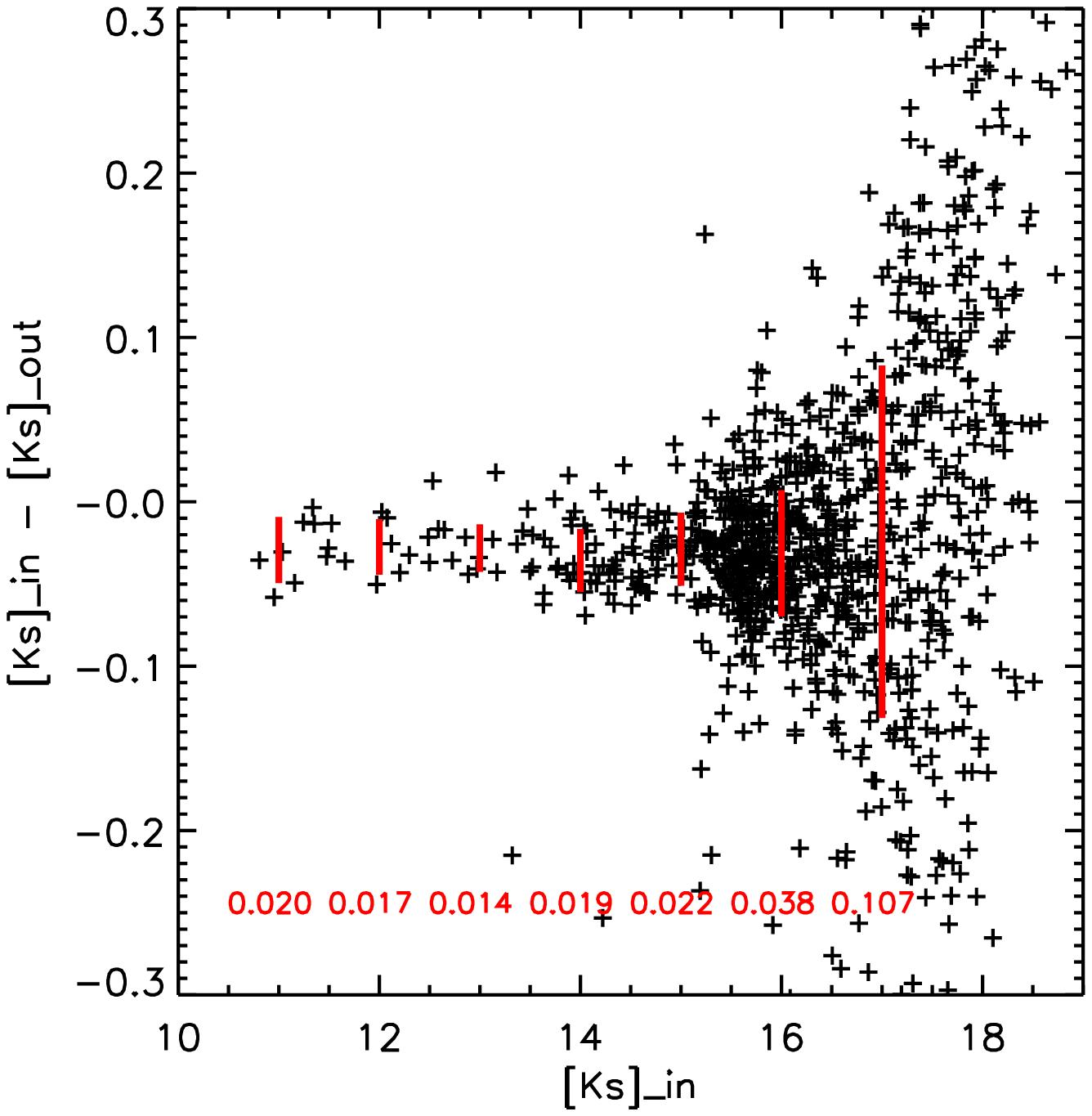}
\caption{\label{Fig:sim} Plot of the difference between input and
  output $Ks$-magnitudes vs. input magnitudes for the simulation
  described in section\,\ref{sec:sys}.}
\end{figure}

We produced simulated data in order to test whether holography may
lead to a systematic bias in photometry or astrometry. As input for
the simulation we took the stellar positions and fluxes measured in a
NaCo/VLT image of the Galactic centre from 28 May 2008. The data are
described in the last line of Table\,1 in \citet{Schodel:2009zr}. The
speckle PSFs were simulated with the speckle simulator of
\citet{Rengaswamy:2010fk}, assuming $0.8''$ seeing in the visual, a
wind speed of 10\,m\,s$^{-1}$, an airmass of $1.2$, and an exposure
time of $0.15$\,s.  The gain and readout noise of NaCo's detector were
used.  Readout and photon noise were added to the 4000 artificial
speckle frames. After holographic reconstruction PSF fitting
photometry and astrometry was performed with {\it
  StarFinder}. Figure\,\ref{Fig:sim} shows a plot of the difference
between the input and output magnitude vs. the input magnitude for
each star. There is a small systematic difference of about
0.03\,mag. This shift is due to the uncertainty of the zero point,
which is caused by the exact choice of the parameters for PSF
extraction (mainly the size of circular mask applied to the PSF). But
there is no apparent trend and the scatter is small at all
magnitudes. Hence, from the simulations there is no evidence for any
systematic photometric or astrometric errors that could be introduced
by the holographic technique.

\subsection{Noise properties \label{subsec:noise}}

Holographic image reconstruction can be considered a deconvolution
technique.  As is well known, deconvolution can introduce artifacts
into images, like the typical ringing around stars in the Wiener
deconvolution or the grainy background and photometric bias on faint
sources in the case of the Lucy-Richardson algorithm. Deconvolution
can alter the noise statistics by introducing covariances between the
pixel \citep[the mentioned effects are discussed, e.g.,
in][]{Schodel:2010hc}. The absence of any significant covariances
between pixels is fundamental for the application of PSF fitting
photometry and astrometry programs, like DAOPHOT
\citep{Stetson:1987nx} or {\it StarFinder} \citep{Diolaiti:2000fk}.

Fortunately, speckle holography is a linear algorithm and a large
number of independent frames are averaged before division in Fourier
space. While graininess of the background (covariances between pixels)
can occur in images reconstructed from a small number of speckle
frames, this effect rapidly disappears with an increasing quantity of
data. 

It is {\it a priori} not clear whether one can estimate the noise in
images reconstructed with the holography method in the same way as in
standard long-exposure images, where readout and photon noise usually
dominate the noise statistics. It is not necessarily trivial, how the
(unavoidable) errors made at estimating the instantaneous PSFs are
translated into the final reconstructed image. Also, particularly at
short integration times, additional sources of uncertainty may become
important, like systematic errors in the applied flat-fields or errors
introduced by correlated noise in the readout electronics (often
visible in short-exposures, e.g., in the form of horizontal
stripes). Furthermore, the signal-to-noise ratio in mosaicked images
will generally not be constant across the image.

\begin{figure}
\includegraphics[width=\columnwidth,angle=0]{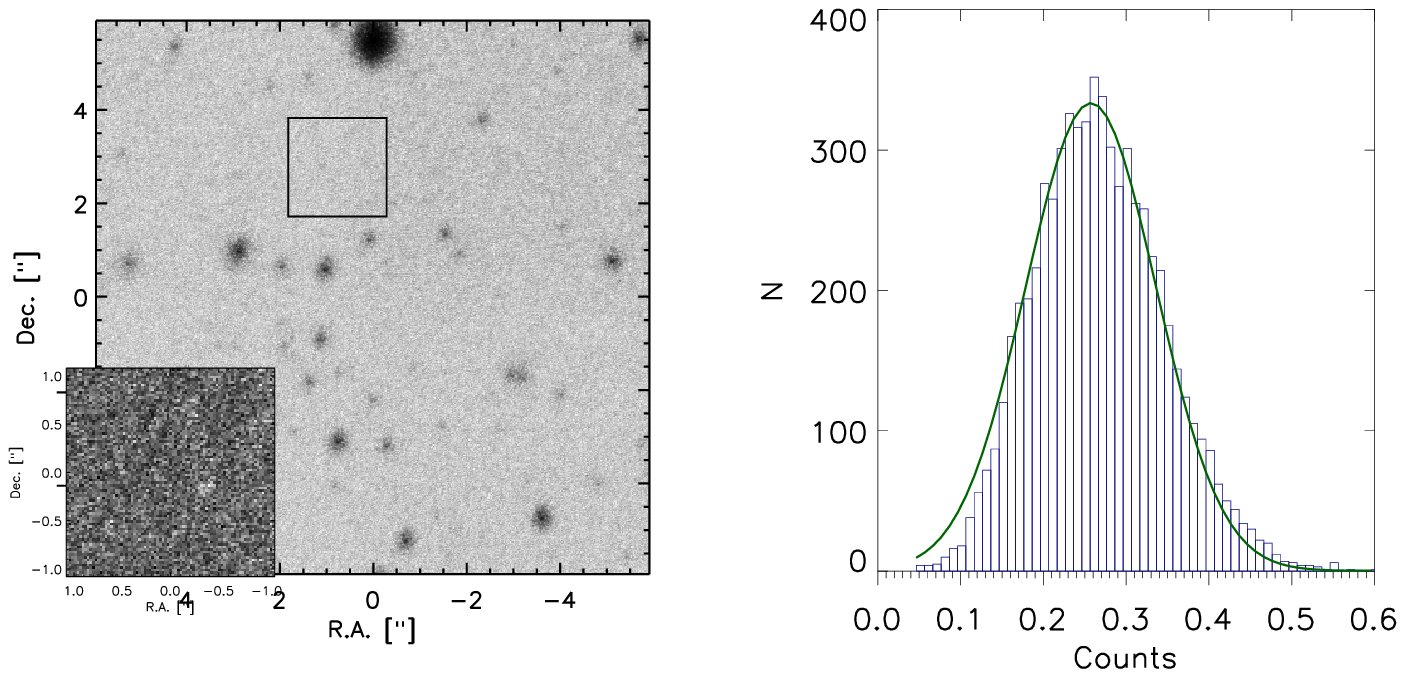}
\caption{\label{Fig:noise} Left: Noise map for the reconstructed
  $Ks$-band image of the GC. The inset shows a zoom into a region of
  the noise map devoid of bright stars. Right: Histogram of pixel
  values in top right image, with Gaussian fit over-plotted. }
\end{figure}

A simple and reliable way of creating noise maps that include
basically all of those effects is through creating images from
independent sub-sets of the data, so-called {\it sub-maps}. The noise
at each pixel can then be estimated as the uncertainty of the mean of
the sub-maps at a given pixel. As an example, in the left panel of
Fig.\,\ref{Fig:noise}, we show the noise map, estimated from 7
sub-maps, for the $Ks$-band image of the GC
(Fig.\,\ref{Fig:reduction}, bottom panel). The inset shows the
magnified view of a region devoid of bright stars. There appears to be
no obvious cross-correlation between the pixels. The auto-correlation
of this region shows no obvious structures, apart from a central peak
at offset $(0,0)$. If the peak of the auto-correlation map is
normalized to one, then all other values are $\leq5\times10^{-5}$. The
histogram of the pixel values in the inset is shown in the the bottom
left panel. The over-plotted fit shows that the pixel values follow
closely a Gaussian distribution.  We inferred from experiments with
the data that Gaussian noise statistics can be obtained already with
only several hundred frames.

\subsection{Estimating uncertainties}

For astrometry and photometry we used {\it StarFinder}, a PSF fitting
program package, that was specifically developed for the analysis of
AO images. Given a noise map for an image, {\it StarFinder} will
calculate the formal uncertainties of the positions and fluxes of
point sources. This noise map can be calculated from the relevant
quantities, like read-out-noise, dark current, gain, number of
exposures. Gaussian noise can also be estimated directly from the
image with the help of the {\it StarFinder} routine
GAUSS\_NOISE\_STD. 

\begin{figure}
\includegraphics[width=\columnwidth,angle=0]{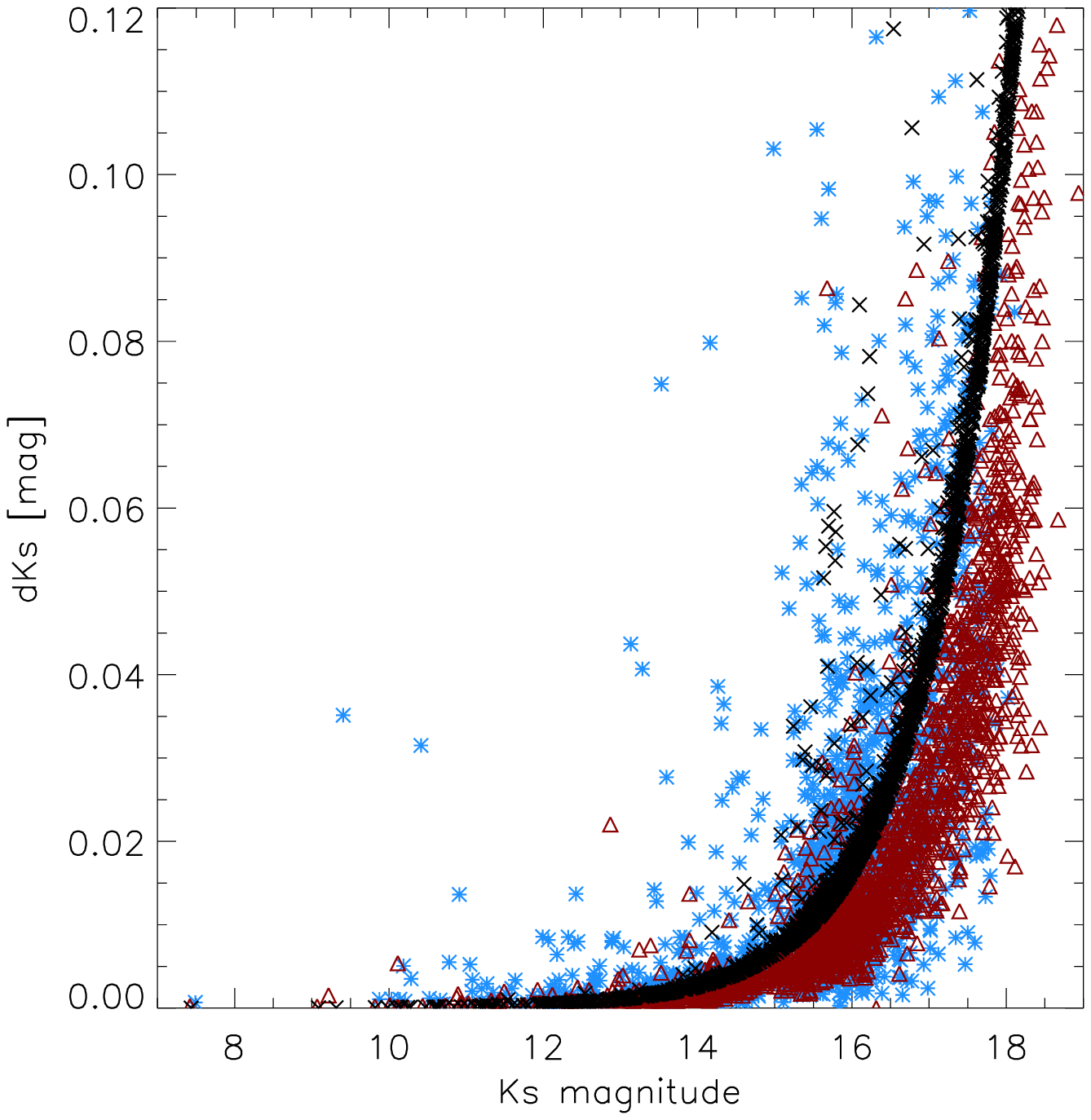}
\caption{\label{Fig:noiseestimate} Photometry on the holographically
  reconstructed $Ks$-band image of the GC: Comparison of different
  methods to estimate the uncertainties. The black crosses, arranged
  in a line with very low scatter, represent the formal uncertainties
  estimated by {\it StarFinder}, using only Gaussian and photon noise
  estimated with {\it StarFinder} routines. The blue asterisks show
  uncertainty estimates from comparing the photometry obtained on
  images reconstructed from three separate sub-sets of the data. The
  red rectangles show the formal uncertainties computed by {\it
    StarFinder} when using a noise map derived from the images
  reconstructed from seven separate sub-sets of the data.}
\end{figure}

We explored three different methods to estimate photometric and
astrometric uncertainties: (1) The {\it formal method}, in which
uncertainties are evaluated by {\it StarFinder}, given a noise map
composed only of Gaussian plus photon noise. The Gaussian noise was
evaluated directly from the image with the GAUSS\_NOISE\_STD routine.
(2) The {\it sub-map method}: Instead of using a noise map, $N$
independent images are analysed. The uncertainties are estimated from
the error of the mean of the stellar positions and fluxes. The
independent images, termed {\it sub-maps}, are created from
non-overlapping subsets of the data, with each set containing a
fraction of $1/N$ of all frames. (3) The {\it noise map method}: The
sub-maps (see previous point) are used to create a noise map, in which
each pixel is assigned the error of the mean at its location (i.e.,
standard deviation divided by $\sqrt{N}$). This noise map is then
input into {\it StarFinder}. Note that estimating the noise is itself
a process that is subject to uncertainties. Therefore, for a small
number of sub-maps, e.g. $N=3$, the noise map may contain pixels that
deviate significantly from the local mean noise. For small $N$, it can
therefore be good practice to median-smooth the noise map, e.g., with
a square box of 3 to 4 pixels width. Here, we use $N=7$ to create
noise maps, with no smoothing.

A comparison of point source photometry with the three methods is
shown in Fig.\,\ref{Fig:noiseestimate} ($N=3$ for the sub-map method
and $N=7$ for the noise map method). There is very good general
agreement between the three methods. It therefore appears that a
holographically reconstructed image can be analysed with PSF fitting
software in the same way as standard long-exposures or AO images.
The formal method shows hardly any scattering of the uncertainties and
very few outliers. This indicates that the true uncertainties may be
significantly under-estimated for some sources.
As concerns the other two methods, visual inspection of the image
revealed that stars with uncertainties significantly above the mean
for a given magnitude, were located close to brighter companions. This
agrees with our expectations. Note that the sub-map method with $N=3$
leads, as expected, to a larger scatter in the uncertainties and to a
reduced sensitivity compared to the noise map method.

In contrast to the formal method, which tends to lead to the detection
of numerous spurious sources in the halos of bright stars, both the
sub-map and the noise-map methods lead to robust source detection, as
long as adequate flux and point-source correlation thresholds are
applied.  The sub-map method has been applied previously to both
speckle and AO data \citep{Ghez:2008fk}. We consider the noise map
method the most attractive technique because it combines robust point
source detection with high sensitivity and realistic estimates of the
uncertainties directly from the data, thus including a broad range of
error sources (photons, readout, electronics, flat field etc.). Note
that the noise map method is -- to the best of our knowledge -- a
simple yet new method that can be attractive for a broad range of
astronomical images, not only for holographically reconstructed ones.
For example, the noise map method can also be applied to AO images and
may serve to estimate the uncertainties in deconvolved images, too.

\subsection{Spatial and temporal stability of the PSF}

\begin{figure}
\includegraphics[width=\columnwidth,angle=0]{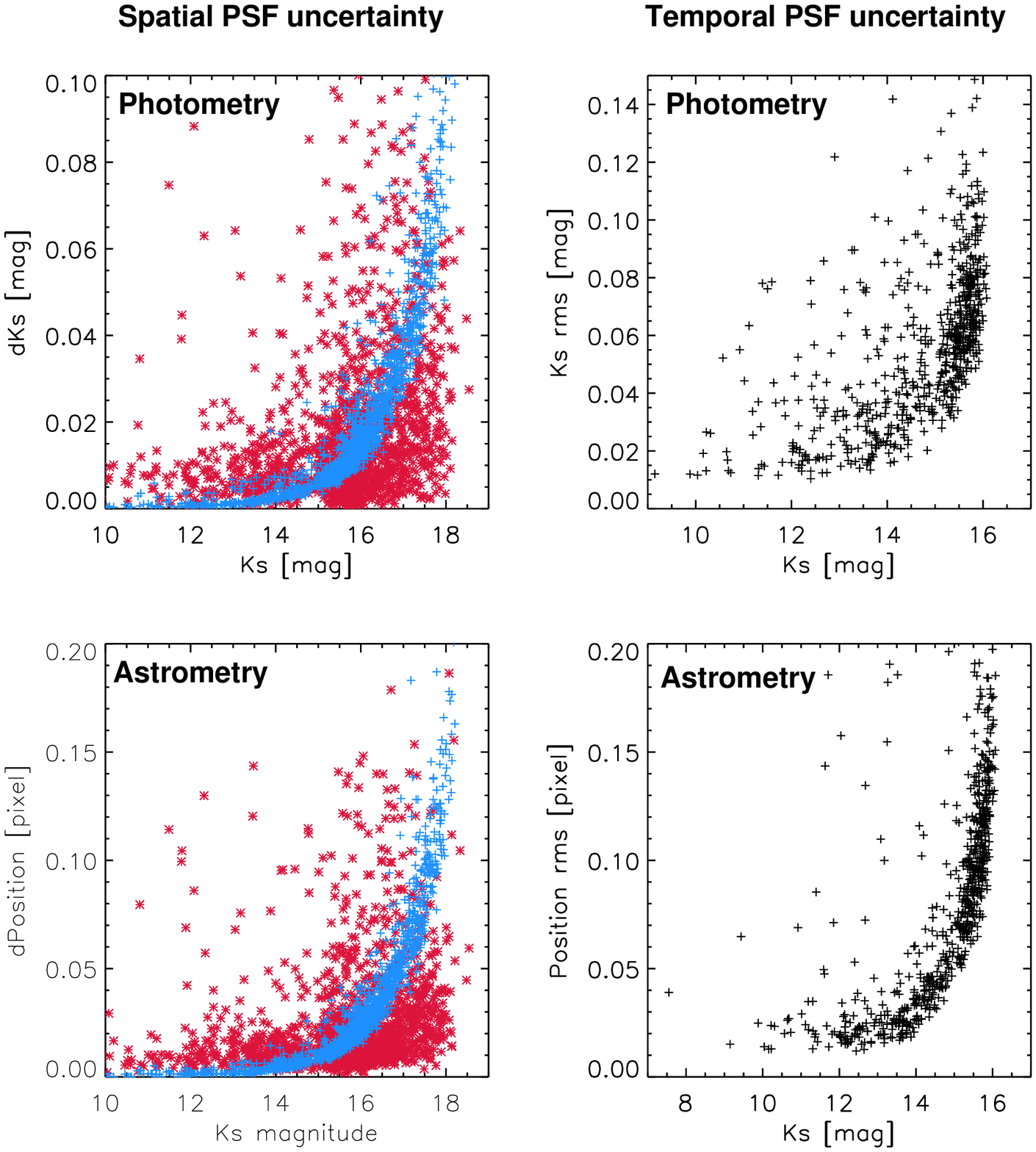}
\caption{\label{Fig:stability} Top left: Photometric uncertainties
  caused by spatial variations of the PSF in the holographic $Ks-$band
  image of the GC. Blue crosses represent the statistical
  uncertainties, red asterisks show the uncertainties caused by the
  spatial variability of the PSF. Top right: Photometric uncertainties
  caused by temporal variations of the PSF for the same data. Bottom
  left and right: As top left and right, but for astrometric
  uncertainties.}
\end{figure}

{\it StarFinder} assumes a perfectly known and spatially stable PSF,
which is, of course, an assumption that will always be violated to a
certain degree, mainly because of anisoplanatic effects. We examined
this problem on the GC $Ks-$band image
(section\,\ref{sec:algorithm}). We performed repeated PSF fitting
photometry and astrometry with three different PSFs, estimated from
independent sets of reference stars. Note that mean offsets in
photometry and astrometry will appear when an image is analyzed with
slightly different PSFs. Those mean offsets (on the order of
$0.02-0.03$\,mag for photometry and $0.001-0.01$\,pixels in each axis
for astrometry) are not relevant for our analysis and were removed
before combining the lists resulting from the different runs of {\it
  StarFinder}. Subsequently, uncertainties were estimated from the
standard deviation of the measurements with the three different
PSFs. The photometric uncertainties are shown in the left panel of
Fig.\,\ref{Fig:stability}. As expected, the PSF uncertainties appear
to be independent of stellar magnitude and are small, with a median of
$0.008$\,magnitudes. The astrometric uncertainty, shown in the right
panel of Fig.\,\ref{Fig:stability}, has a median uncertainty of
$<0.01$\,pixels in each axis. Note that PSF uncertainties are the
dominating source of uncertainty for sources of $Ks\leq15$.

The photometric and astrometric uncertainties from the {\it temporal}
fluctuation of the PSF were estimated by running {\it StarFinder} on
images reconstructed from 25 consecutive data cubes of $\sim$$500$
frames each (GC $Ks-$band data).  The uncertainties caused by temporal
fluctuations of the PSF were estimated from the standard deviation of
the measurements on the 25 different images, after correction of the
mean offsets (which were somewhat smaller than for the spatial PSF
variation). They are shown in the right panel of
Fig.\,\ref{Fig:stability} and dominate the uncertainties for bright
stars. They are $\leq0.02$\,mag for stars brighter than
$Ks\approx14$. The corresponding astrometric uncertainties are
$\leq0.03$\,pixel in each axis for stars brighter than $Ks\approx14$.

We conclude that holographic imaging can lead to images with a
spatially and temporally stable PSF. The photometric uncertainty -
using the noise map method preferred by us - is $\leq0.01$\,mag for
$Ks\leq15$ stars. The $3\sigma$ detection limit in the holographically
reconstructed image of the GC is $Ks\approx19$, for a total
integration time of 1875\,s.

\section{$K-$band imaging of the Galactic centre with Keck/NIRC}
\label{sec:Keck}

Between 1995 and 2005 the Galactic centre was observed with the
speckle camera NIRC at the Keck telescope in order to monitor the
motions of stars around Sgr\,A* \citep[e.g.][]{Ghez:2008fk}. We have
applied our holographic image reconstruction method to these data.
They are of special interest because the holographic technique allows
us to detect fainter sources than in the previously used SSA images.
Thus, it becomes possible to track the orbital motion of stars that
are too faint to have been picked up in the previously used SSA
images. Also, crowding can lead to systematic errors in the measured
positions of the stars close to Sgr\,A*
\citep[see][]{Ghez:2008fk,Gillessen:2009qe}. The excellent Strehl
ratio of the holographic images enables us to better disentangle close
sources than in previous analyses of these data. Since the methodology
presented in this paper was in large part developed and fine-tuned
with NIRC speckle data of the GC and since the holographically
reconstructed NIRC/Keck GC data have been used in recent work
\citep[][]{Meyer:2012fk} and will be used in future publications, this
work is the natural place to explain the corresponding details.

As an example, we present a data set from April 2005. About $10,000$
speckle frames with an exposure time of $0.1$\,s were used for image
reconstruction. The stars IRS\,16C, IRS\,16NW, and IRS\,16SW were used
as point source references \citep[$Ks=9.9$, $Ks=10.1$, and
$Ks\approx10.2$, see][]{Schodel:2010fk}.

A special feature of the NIRC GC speckle data is that they were
collected in the so-called {\it stationary mode}, which keeps the
pupil fixed on the detector. This avoids a variable contribution of
the secondary spider to the diffraction patterns. A part of the FOV
was masked in all NIRC data, with the mask changing relative to the
stars because of the sky rotation. In order to deal with the variable
FOV, we constructed a weight map to adjust the final reconstructed
image. The weight map was created by running the corresponding masks
of all the speckle frames through the same holographic reduction
pipeline.  We used an Airy PSF for a circular 10m-aperture for
apodisation of the final image. The FWHM of this Airy function,
$\sim$$0.05"$, corresponds to the angular size of a resolution
element of the Keck telescopes in the $K-$band.

\begin{figure}
\includegraphics[width=\columnwidth,angle=0]{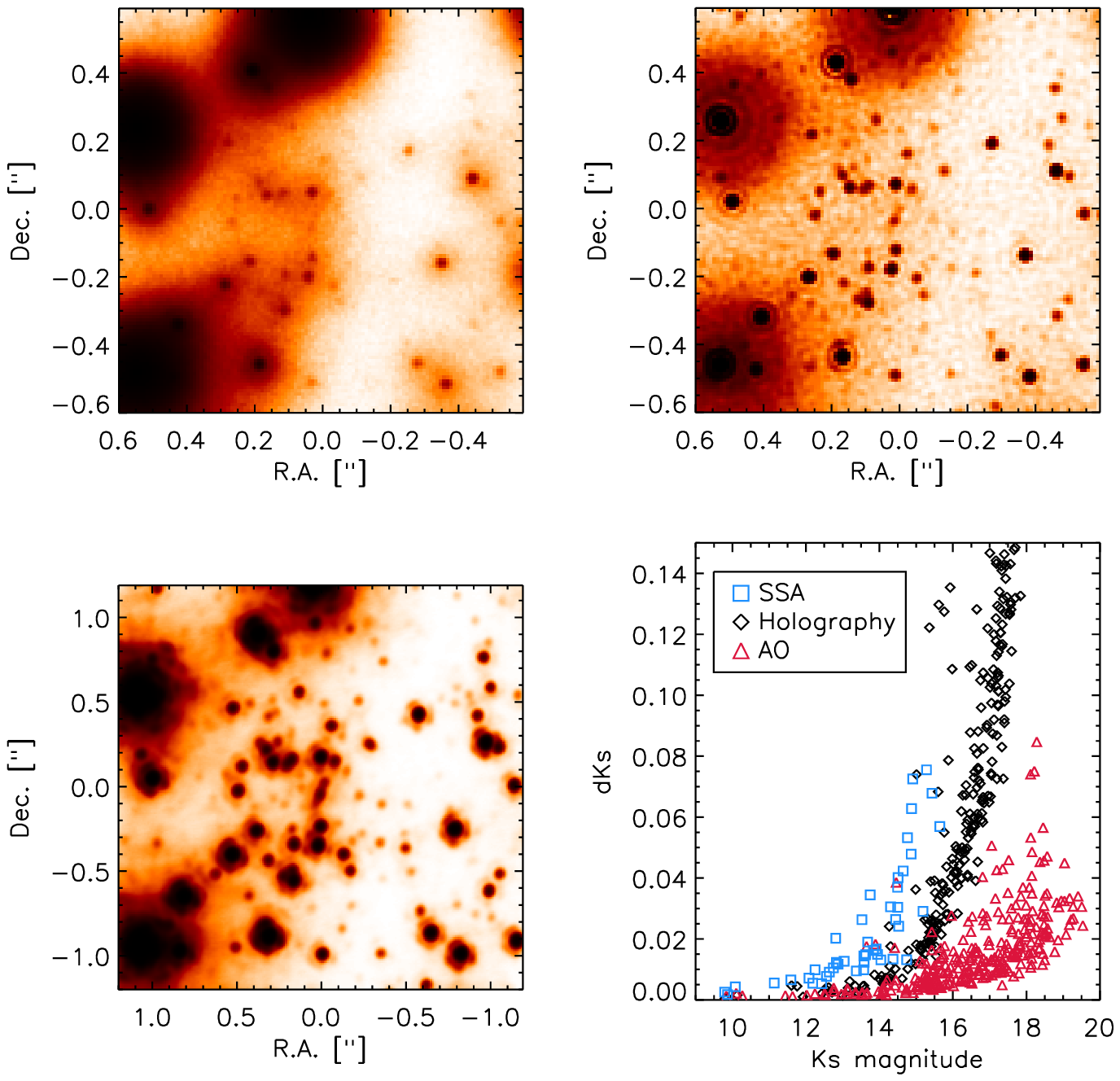}
\caption{\label{Fig:GC} Speckle holography of the Galactic centre with
  NIRC/Keck. Upper left: SSA image of the central arcseconds of the
  Galactic centre  from April 2005.  Upper right: Image
  reconstructed with the holography technique. Lower left: NIRC2/Keck
  LGS AO image of the Galactic centre from July 2005. Lower right:
  Photometric uncertainty vs.\ $K$-magnitude for sources detected
  within $2"$ of Sgr\,A*: blue squares: SSA; black diamonds:
  holography; red triangles: AO. The detection threshold was set to
  $5\,\sigma$.}
\end{figure}

In Fig.\,\ref{Fig:GC} we show close-ups on the environment of Sgr\,A*
from an SSA and a holography image for the April 2005 data, as well as
from a laser guide star assisted AO image from July 2005
\citep{Ghez:2008fk}.  A plot of photometric uncertainty vs. magnitude
for the point sources detected in the three images is shown in the
same Figure. Lucky imaging was applied to the SSA reconstruction, with
selection of the best 10\% of the frames. The exposure times of the AO
($\sim$900\,s) and holography ($\sim$1000\,s) images are
comparable. As can be seen in Fig.\,\ref{Fig:GC}, holographic image
reconstruction provides excellent spatial resolution with a very well
calibrated PSF in the reconstructed image. The Strehl is about
$40\%$. As shown in the bottom right panel of Fig.\,\ref{Fig:GC}, the
holography image is roughly 2\,mag deeper than the SSA image, while
the AO image is about 2\,mag deeper than the holography image.

\label{lastpage}

\end{document}